\documentclass[twocolumn,aps,prb,showkeys,amsmath,amssymb]{revtex4}
\usepackage{graphicx}
\usepackage{dcolumn}
\usepackage{bm}
\usepackage{epsf}

\newcommand{{\vergul}}{  ,}

\begin{document}
\title{Retroreflector Approximation of a Generalized Eaton Lens}
\author{Sang-Hoon  \surname{Kim}} \email{shkim@mmu.ac.kr }

\affiliation{
Division of Marine Engineering,
Mokpo National Maritime University,\\
Mokpo 530-729, R. O. Korea}
\date{\today}


\begin{abstract}
We extended a previous study of the Eaton lens at specific refraction angles
to the Eaton lens at any refraction angle.
The refractive index of the Eaton lens is complicated
and has not analytical form  except at a few specific angles.
 We derived a more accessible form  of the refractive index for any refraction angle
 with some accuracy by retroreflector approximation.
The finding of this study will be useful for a rapid estimation of the refractive index,
and the the design of various Eaton lenses.
\end{abstract}

\maketitle

\section{Introduction}

Trajectories of light can be controlled by prisms or a
combination of mirrors, as well as  by
controlling the refractive index(RI) of a lens, such as
 GRIN(Gradient Index) lenses
like the Eaton lens, Luneburg lens, and Maxwell's fish-eye lens, etc.\cite{smoly}
It was once thought that the control of light trajectories was unrealistic
or very difficult to realize, but  recent developments of
transformation optics and high RI materials from metamaterial
techniques have provided novel methods for controlling wave trajectories.

The Eaton lens is a typical GRIN lens in which the RI varies from one to
infinity.
It has a singularity in that the RI goes to
infinity at the center of the lens
and it originates from a peculiar dielectric.
The speed of light is reduced to zero at this point, and the lens can therefore,
 change the wave trajectories any direction.

The Eaton lens was recently studied at three specific refraction angles:
$90^o$(right-bender), $180^o$(retroreflector), and
$360^o$(time-delayer).\cite{hannay,tomas,danner}
The RI  of the Eaton lens is given as a function of the radius, but
it is not analytic except for at a few specific angles.
Generally, the RI and its trajectories can only be obtained by numerical calculations.
A simple but good approximation of the general form
will be helpful for the design and application of Eaton lenses.

The RI of the Eaton lens is extended to arbitrary refraction angles.
By combining a few Eaton lenses, we can construct an optical triangle, square,
 hexagon, or any  geometric shaped path without mirrors.
An extremely  simple form of the RI is
suggested within a reliable error range for easy and practical use
by utilizing a linear approximation of the retroreflector.

\section{Generalized Eaton Lens}

Hendi et al.\cite{hendi} performed studies of light trajectories
at central potentials.
Based on  the fact that the trajectory of the
Eaton lens is an analogue of Kepler's scattering problem
 ({\it ``scattering tomography"}),
they derived the relation between the light trajectory
and its RI at any impact parameter.
For symmetric and spherical lenses,
the RI at the three specific refraction angles described
above are known as\cite{danner}

\begin{figure}[htb]
\centerline{\includegraphics[width=7.5cm]{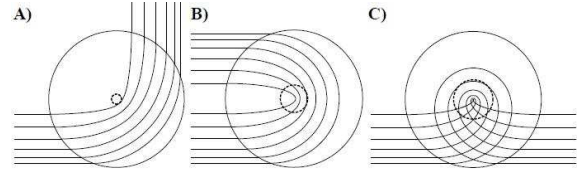}}
\caption{\label{fig:eaton1} The trajectories of Eaton lenses at
three specific angles from Ref. \cite{danner}.
(A)$90^o$, (B)$180^o$, (C)$360^o$.
 Reprinted with permission from OSA.}
\end{figure}

\begin{align}
 n^2 &=&
\frac{1}{nr}+\sqrt{\frac{1}{n^2 r^2}-1} \hspace{10pt} &(\theta = 90^o),&
\label{10} \\
 n &=& \sqrt{\frac{2}{r}-1} \hspace{50pt} &(\theta = 180^o),&
\label{12} \\
 \sqrt{n} &=& \frac{1}{nr}+\sqrt{\frac{1}{n^2 r^2}-1}  \hspace{9pt} &(\theta = 360^o).&
\label{14}
\end{align}
$n=1$ for $r \ge 1$, where  $n$ is the relative RI.
 The $r$ is the radial position between 0 and 1, and
the actual radial position is given by $a r$, where $a$ is the
radius of the lens.

The importance of the above three lenses has been discussed previously.
 The right-bender in Eq. (\ref{10})  has been studied
 in  bending surface plasmon polaritons.\cite{zent}
 The retroreflector in Eq. (\ref{12}) is a device
 that returns the incident wave back to its source.\cite{ma}
The time-delayer in Eq. (\ref{14}) is a time-clocking device in which
the incident wave exits in the direction of the incidence as if the
lens were not present. The trajectories for the above three cases
was plotted by Danner and Leonhardt\cite{danner}
and is shown in Fig.~\ref{fig:eaton1}.

\begin{figure}[htb]
\centerline{\includegraphics[width=7.5cm]{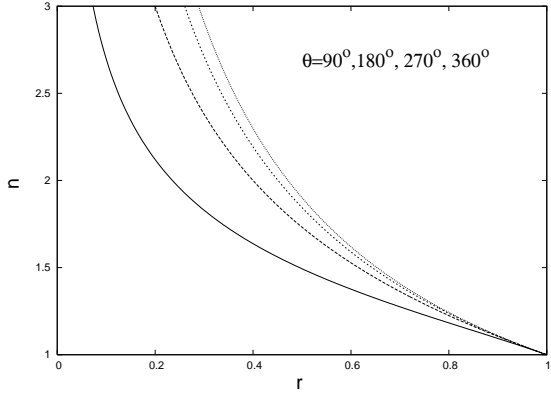}}
\caption{\label{fig:eaton2} The refractive indexes at $\theta=90^o,
180^o, 270^o$, and $360^o$ from left to right.}
\end{figure}

\begin{figure}[htb]
\centerline{\includegraphics[width=7.5cm]{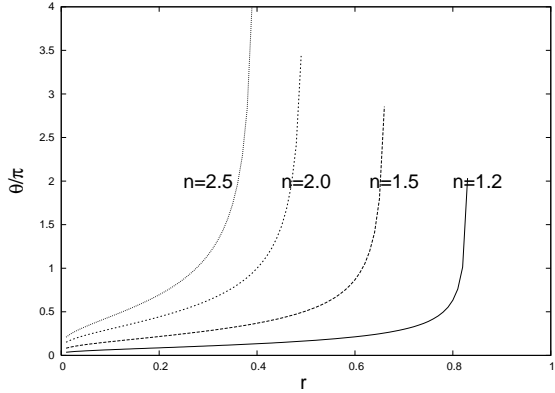}}
\caption{\label{fig:eaton3}
 The refraction angles at various refractive index and radius.}
\end{figure}

The RI of arbitrary refraction angles was derived from
 optics-mechanics analogy.
  RI is identical to particle trajectories of equal
total energy $E$ in a central potential $U(r)$.
The potential of the Eaton lens corresponds to the gravitation-like
$U \propto -1/r$.\cite{hannay}
On the other hand the potential of the Luneburg lens
corresponds to the harmonic-like $U \propto r^2$.\cite{morgan}
From the conservation of mechanical energy
inside and outside the lens, the kinetic energy of a particle in the
medium of RI $n$
 is written as $(1/2)m n^2 v^2 = E-U$, where $m$ is the mass
  and $v$ is the velocity of the particle inside the lens.
Then, $\int n ds$ or $\int \sqrt{E-U}ds$ should be stationary bases on Fermat's principle.

Replacing $\chi +\pi=\theta_{final}-\theta_{initial}$ into $\theta$
of Hannay and Haeusser's equation\cite{hannay}
(See reference for the detailed derivation),
we obtain a generalized form of the RI of the
Eaton lens for arbitrary refraction angles as
\begin{equation}
 n^{\pi/\theta} = \frac{1}{nr}+\sqrt{\frac{1}{n^2 r^2}-1},
\label{20}
\end{equation}
where $\theta$ is any radian angle and
 $n=1$ for $r \ge 1$.
The refraction angle can be generalized as $\theta=(2N+1/2)\pi$ for
the right-bender, $\theta=(2N+1)\pi$ for the  retroreflector, and
 $\theta=2N\pi$ for the time-delayer, where $N$ is an integer.
 Note that $\theta = \pm \pi$ produces the same result with Eq. (\ref{12}).
We calculated RI numerically and plotted
it at four specific angles  in Fig.~\ref{fig:eaton2}

Using the logarithm provides a more convenient form of the relation
between RI and the radius.
\begin{equation}
\frac{\theta(n,r)}{\pi} = \frac{\log
n} {\log \left( \frac{1}{nr}+\sqrt{\frac{1}{n^2 r^2}-1} \right)}.
\label{30}
\end{equation}
These relations are plotted in Fig.~\ref{fig:eaton3}.
A large RI changes trajectories at a small radius.
The relation between RI and the radius is relatively more convenient to find
 using Eq. (\ref{30}) than Eq. (\ref{20}),
but is still not  practical for making a rapid estimation.

\section{Retroreflector Approximation}

\begin{figure}[htb]
\centerline{\includegraphics[width=7.5cm]{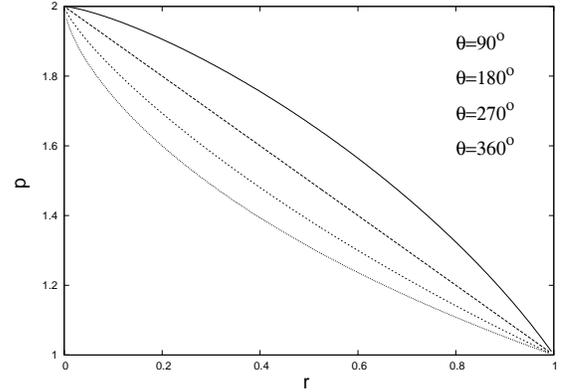}}
\caption{\label{fig:eaton4} $p(r)$ as a function of radius.
$\theta=90^o, 180^o, 270^o$, and $360^o$ from top to bottom. $p(r)$
is linear for $\theta=180^o$.}
\end{figure}

Introducing a new radial function $p(r)$,
the nonlinear Eqs. (\ref{20}) and (\ref{30}) are used as
a more accessible form.
From Eq. (\ref{20}) we obtain $n(r)$ at two boundaries. As $r \sim
0$,  then $n \sim (2/r)^{\theta/(\pi + \theta)}$.
As $r \sim 1$,
then $n \sim (1/r)^{\theta/(\pi + \theta)}$. Therefore, we can write
$n(r)$ for the range of $0< r \le 1$ in the following compact form
\begin{equation}
 n(r) = \left\{ \frac{p(r)}{r}\right\}^{\frac{\theta}{\pi+\theta}},
\label{40}
\end{equation}
where $0 < p(r) \le 1$.
The new variable $p(r)$ is plotted in Fig.~\ref{fig:eaton4}.

The time-delayer with $\theta=2N\pi$ is an invisible sphere. The
general form with an $N$ turn  can be represented easily using $p(r)$
in Eq. (\ref{40}). When it has an $N$ turn,
the time delay $\triangle \tau$ is obtained as
\begin{equation}
\triangle \tau = \frac{2N\pi r n}{v_o}
=\frac{2N\pi r }{v_o}\left\{ \frac{ap(r)}{r} \right\}^{\frac{2N}{2N+1}},
\label{55}
\end{equation}
where $v_o$ is the background velocity outside the
lens and $a$ is the radius of the lens. If $N \gg 1$, then $2 N \pi
a/v_o \le \triangle t \le 4 N \pi a/v_o$.
Therefore, $a$ and $N$ are
the two main factors that decide the time cloaking or delay properties.

As  $p(r)$ is a function of the radius, we can introduce an approximation.
From Eq. (\ref{40})  $p(r)$ can be written as
\begin{equation}
 p(r) = r n(r)^{\frac{\pi+\theta}{\theta}}.
 \label{45}
 \end{equation}
It is a bounded and analytic function of the radius between 1 and 2.
Every line is monotonically decreasing from $r=0$ to $r=1$ and nearly linear.
Therefore, we can take the retroreflector approximation
or a linear approximation as
\begin{equation}
p(r)\simeq 2-r.
\label{48}
\end{equation}
Substituting Eq. (\ref{48}) into Eq. (\ref{40}),
we obtain the simple form of the RI at any refraction angle as
\begin{equation}
n(r) \simeq
\left( \frac{2}{r} -1\right)^{\frac{\theta}{\pi+\theta}}.
\label{50}
\end{equation}

\begin{figure}
\centerline{\includegraphics[width=7.5cm]{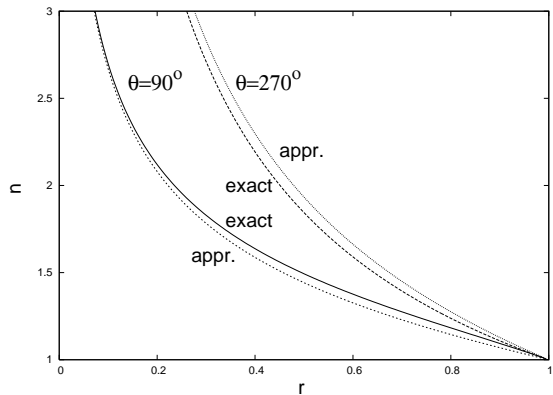}}
\caption{\label{fig:eaton5} Comparison between the exact value
obtained by numerical calculation and the retroreflector
approximation for $\theta=90^o, 270^o$.
The error is less than 4$\%$ at $\theta=90^o$, and less than 5 $\%$
 at $\theta=270^o$. They match exactly
 at the both ends of $r \rightarrow 0, 1$ at every angle. }
\end{figure}
\begin{figure}
\centerline{\includegraphics[width=5.5cm]{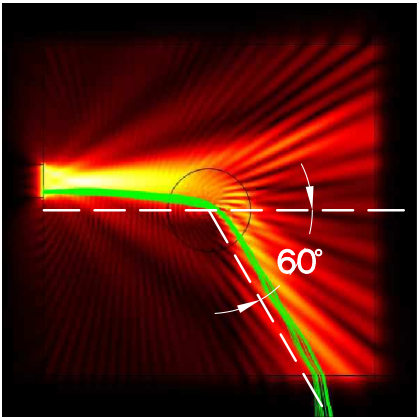}}
\caption{\label{fig:eaton6} Optical path at $\theta=60^o$ by
the retroreflector approximation in Eq. (\ref{50}).
The plot was generated using a COMSOL simulator.}
\end{figure}

The effectiveness of the approximation is examined by comparison
with the exact values obtained from the numerical calculations in
Fig.~\ref{fig:eaton5} and
the optical path was plotted for $\theta=60^o$ in Fig.~\ref{fig:eaton6}.
The error of Eq. (\ref{50}) is less than 4$\%$ at $\theta=90^o$,
5$\%$ at $\theta=270^o$, and 10$\%$  at $\theta=360^o$.
 The error accumulates for more than $360^o$ and less than $0^o$.
 Therefore, this calculation is useful at the conventional angle ranges of
 $0< \theta <2\pi$ within a 10$\%$ error.

\section{Conclusions}

The Eaton lens is a typical GRIN lens with a complicated RI.
We extended previous studies of the Eaton lens at specific refraction angles
to any refraction angle.
The RI refractive index of the Eaton lens is complicated and not analytical
 except for a few specific angles.
 We derived a more accessible form of the RI
for any refraction angles using the retroreflector approximation.
This method is applicable at conventional angle ranges
of $0 < \theta < 2\pi$ within a $10\%$ error.
It is extremely simple but useful for a rapid estimation of optical properties
obtained from the RI.

\section*{Acknowledgments}
The author would like to thank  S. H. Lee and M. P. Das
for useful discussions.
This research was supported by Basic Science Research Program through
the National Research Foundation of Korea(NRF)
funded by the Ministry of Education, Science and Technology(2011-0009119)


\begin{thebibliography}{0}

\bibitem{smoly} V. N. Smolyaninova, I. I. Smolyaninov, A. V. Kildishev, and V. M. Shalaev, Maxwell fish-eye and Eaton lenses emulated by microdroplets,
      {\it Opt. Lett.} {\bf 35} (2010) 3396.
\bibitem{hannay} J. H. Hannay and T. M. Haeusser, Retroreflection by refraction,
{\it J. Mod. Opt.}  {\bf 40} (1993) 1437.
 \bibitem{tomas} T. Tyc and U. Leonhardt, Transmutation of singularities in optical instruments, {\it N. J. Phys.}  {\bf 10} (2008) 115038.
\bibitem{danner} A. J. Danner and U. Leonhardt, Lossless design of an Eaton lens
and invisible sphere by transformation optics with no bandwidth limitation,
2009 Conference on Lasers and Electro-Optics(CLEO), Baltimore, MD,
U. S. A. (2009).
\bibitem{hendi} A. Hendi, J. Henna, and U. Leonhardt, Ambiguities in the Scattering Tomography for Central Potentials,  {\it Phy. Rev. Lett.} {\bf 97} (2006) 073902.
\bibitem{zent} T. Zentgraf, Y. Liu, M. H. Mikkelson, J. Valentine, and X. Zhang,
Plasmonic Luneburg and Eaton lenses, {\it Nature, Nano}, {\bf 6} (2011) 151.
\bibitem{ma} Y. G. Ma, C. K. Ong, T. Tyc, and U. Leonhardt,
An omnidirectional retroreflector based on the transmutation of dielectric singularities, {\it Nature, Mat.} {\bf  8}  (2009) 639.
\bibitem{morgan} S. P. Morgan, General Solution of the Luneberg Lens Problem,
{\it  J. Appl. Phys.} {\bf 29} (1958) 1358.

\end{thebibliography}
\end{document}